\documentclass{article}
\usepackage{spconf,amsmath}
\usepackage[utf8]{inputenc} 
\usepackage[T1]{fontenc}    
\usepackage{hyperref}       
\usepackage{url}            
\usepackage{booktabs}       
\usepackage{amsfonts}       
\usepackage{nicefrac}       
\usepackage{microtype}      
\usepackage{xcolor}         
\usepackage{float}
\usepackage[thinlines]{easytable}
\usepackage{subfig}
\usepackage[export]{adjustbox}
\usepackage{lipsum}
\usepackage{graphicx}
\usepackage{pdflscape}

\title{Noise Perturbation for Saliency Prediction with \\ Psychophysical Synthetic Images}
\name{Qiang Li\thanks{$^*$ Current Address \\
Corresponding Author: qiang.li@uv.es}}
\address{Image Processing Lab, University of Valencia, Spain \\
$^*$ Tri-Institutional Center for Translational Research in Neuroimaging and Data Science (TReNDS) \\
Georgia State University, Georgia Institute of Technology, Emory University, Atlanta, GA, USA}
\begin{document}
%
\maketitle
\begin{abstract}
Convolutional neural networks (CNNs) have enjoyed significant success in the realm of natural image saliency prediction. This study's principal objective is to assess the performance of saliency prediction models based on both CNNs and classical models when applied to psychophysical synthetic images subjected to noise perturbation. We aim to determine whether the performance of these models remains as robust as when they are applied to natural images. Simultaneously, we seek to explore the connection between CNNs and human vision, particularly with regard to low-level vision functions. An overarching question we address is whether CNNs can be considered faithful replicas of human visual functions. In this investigation, we employ CNNs, Fourier-based models, and spectral models that draw inspiration from low-level vision systems to analyze saliency prediction, focusing on psychophysical synthetic images rather than natural ones. Our findings indicate that saliency prediction models inspired by Fourier and spectral theories surpass currently available pre-trained deep neural networks when applied to psychophysical images subjected to noise perturbations. However, it is noteworthy that these psychophysical models exhibited a higher degree of instability in the presence of noise compared to pre-trained deep neural networks. Additionally, we propose that the examination of CNNs through psychophysical methods holds potential benefits for both visual neuroscience and artificial neural network research.
\end{abstract}
\section{Introduction}
\label{sec:intro}
The concept of the attention function was initially introduced within artificial neural networks and has yielded remarkable success in various computer vision tasks, including image classification \cite{Ashish17} and few-shot learning \cite{PengHZ18}, among others. It is indisputable that Convolutional Neural Networks (CNNs) have achieved significant prominence in both academic research and industrial applications. Furthermore, the Vision Transformer (ViT) has emerged as a highly competitive alternative to state-of-the-art convolutional networks, boasting excellent results while demanding significantly fewer computational resources for training \cite{Ashish17}. The overarching message conveyed by these developments can be distilled into the aphorism, "Attention is All you Need." It is worth noting that while there exist some debates regarding the role of attention in artificial neural networks, delving into this discourse falls beyond the purview of this paper.

In the natural world, an abundance of redundant information exists, and the human visual system encounters challenges in processing this wealth of data due to the inherent information bottleneck within the visual system. Nonetheless, research on attention using physiological methods has a long-standing history, with a particular emphasis on the visual system \cite{Carrasco11,QiangNN22}. According to the three-stage vision information processing model posited by \cite{Zhaoping06}, the initial input information is encoded within the retina at a rate of 20 megabytes per second \cite{Kelly62} and subsequently undergoes a selection phase. This is the locus of attention, where irrelevant information is filtered out. Following the selection stage, information is compressed to a rate of 1 megabyte per second before being relayed to the primary visual cortex (V1), ultimately leaving a mere 40 bits per second of information within V1 \cite{Zhaoping06,George56}. A fundamental and classical principle that underpins the concept of visual attention is the "Efficient Coding Principle" \cite{Barlow61}. Subsequently, a rich body of psychophysical experiments has been conducted to investigate visual attention. Concurrently, various computationally designed attention models have been proposed, drawing inspiration from physiological experiments, such as the pioneering Itti models for visual saliency prediction \cite{Itti98}.

\begin{figure}[htb]
\begin{minipage}[b]{1.0\linewidth}
  \centering
  \centerline{\includegraphics[width=9cm, height=22em]{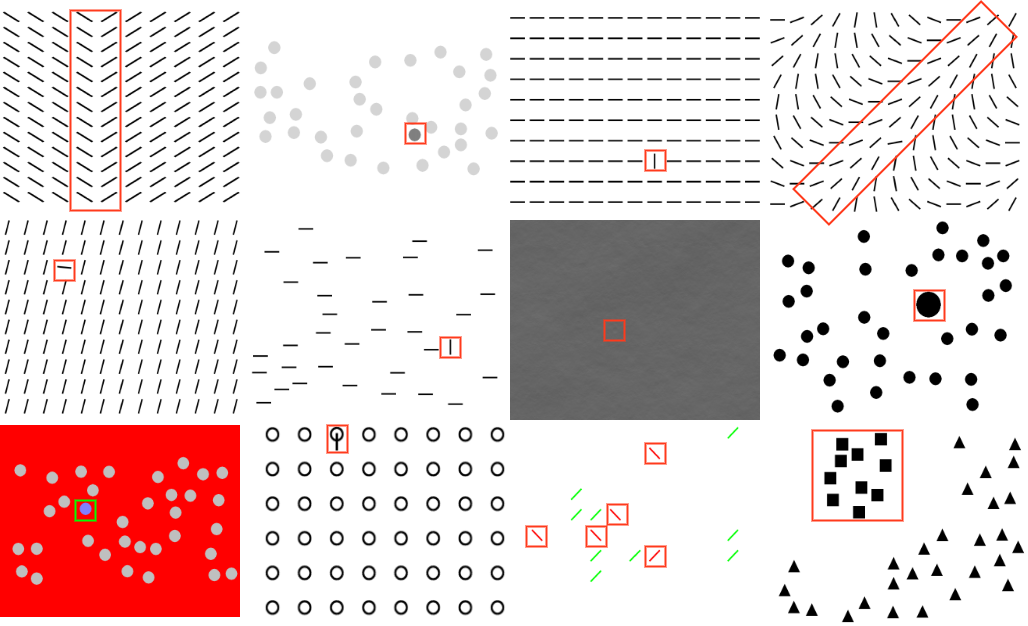}}
\end{minipage}
\caption{\textbf{A subset of images was chosen from the SID4VAM dataset.} The visual conspicuity targets are indicated by a red rectangle in the chosen synthetic images.}
\label{fig:sid4vam_imgs}
\end{figure}

In the age of deep learning, natural image saliency prediction has done very well \cite{Borji2021SaliencyPI}, as shown by ML\_Net \cite{Cornia16}, and DeepGaze II \cite{Kmmerer2016}. However, there are fewer studies that examine saliency prediction on psychophysical images under noise perturbation using CNNs and custom-designed psychophysical models and the relationship between CNNs and psychophysical saliency prediction models. This study is limited to predicting saliency for psychophysically oriented images with consideration of noise rather than natural images. In the pursuit of advancing our understanding in this field, it is imperative to delve into the examination of artificial deep networks alongside the intricacies of the human visual system, with a specific focus on the fundamental aspects of low-level vision functions. At present, the precise nature of the connections that underlie the relationship between Convolutional Neural Networks (CNNs) and low-level vision functions in the human perceptual apparatus remains shrouded in ambiguity. This enigma arises from the perplexing realization that CNNs, as computational constructs, are fundamentally reliant upon the emulation of the very same low-level visual processes inherent to human perception. 

\section{Related Work}
\label{sec:relw}

In the era of deep learning, the field of general saliency prediction on natural images has achieved remarkable success, as highlighted in the work of Borji et al. \cite{Borji2021SaliencyPI}. Predicting visual saliency has emerged as a prominent subject of interest within the domains of machine vision and visual neuroscience. One effective approach to gaining insights into the inner workings of black-box Convolutional Neural Networks (CNNs) is through visualization, with attention maps serving as a valuable tool for comprehending the information processing features inherent in CNNs.

In our investigation, we leveraged classical CNN architectures with varying depths to explore their attention maps. These architectures included AlexNet \cite{NIPS2012_4824}, GoogLeNet \cite{szegedy2014going}, Inceptionv3 \cite{incep}, and Densenet201 \cite{Huang2017DenselyCC}, among others. Additionally, we incorporated several top-performing CNNs explicitly designed for the task of natural image saliency prediction, exemplified by DeepGazeII, which achieved high rankings based on the MIT/Tuebingen Saliency Benchmark evaluation\footnote{https://saliency.tuebingen.ai/}.

Furthermore, our study introduced three of the highest-ranked psychophysical saliency prediction models, each meticulously crafted to align with the functions of the human low-level vision system \cite{QiangLI20}. The first model, known as the HTF model, adopts a bottom-up approach for visual saliency prediction, centering on the frequency domain \cite{Jian12}. The second model, Incremental Coding Length (ICL), was designed based on information-theoretic principles \cite{Hou08}. This model posits that attention regions within images may induce entropy gain in the perceptual state, subsequently attracting high energy. The final model, DCTS, is a spectral-oriented saliency prediction model that forecasts image saliency based on image signatures \cite{Hou11}.

In the ensuing section, we will delve into the process of saliency prediction using all of the aforementioned models when applied to psychophysical images.

\section{Methods}
\label{sec:meth}
\subsection{Dataset: Psychophysical images}
\label{ssec:data}
In this study, we employed the SID4VAM dataset, which comprises 230 synthetic images characterized by well-defined salient regions (refer to Fig.~\ref{fig:sid4vam_imgs}). These images were generated to encompass a total of 15 distinct low-level features, including attributes such as orientation, brightness, color, and size. The generation process involved the creation of synthetic patterns with a target-distractor pop-out configuration. Two different sets of instructions, namely free-viewing and visual search tasks, were used in conjunction with seven feature contrasts for each of the feature categories \cite{Berga2019SID4VAMAB, BERGA201960}.

The pretrained networks and psychophysical models, along with some of their key properties such as depth, parameters, and image input size, are as follows: \\
\\
\textbf{\textit{A. Pretrained Deep Neural Networks:}}
\begin{enumerate}
    \item AlexNet 
    \begin{itemize}
        \item Depth: Deep architecture with multiple convolutional and fully connected layers.
        \item Parameters: Approximately 61 million.
        \item Image Input Size: 227x227 pixels.
    \end{itemize}
    
    \item GoogLeNet
    \begin{itemize}
    \item Depth: A complex architecture with deep and wide inception modules.
    \item Parameters: Roughly 7 million.
    \item Image Input Size: 224x224 pixels.
    \end{itemize}
    
    \item Inceptionv3
    \begin{itemize}
    \item Depth: A deep architecture with inception modules.
    \item Parameters: Around 23 million.
    \item Image Input Size: 299x299 pixels.
    \end{itemize}

    \item Densenet201
    \begin{itemize}
    \item Depth: A densely connected deep neural network.
    \item Parameters: Approximately 20 million.
    \item Image Input Size: 224x224 pixels.
    \end{itemize}

    \item DeepGazeII
    \begin{itemize}
    \item Depth: VGG19 plus several readout networks
    \item Parameters: N/A.
    \item Image Input Size: 224x224 pixels.
    \end{itemize}
\end{enumerate}

\textbf{\textit{B. Psychophysical Models:}}

\begin{enumerate}
    \item HTF Model (Frequency-based) 
    \begin{itemize}
        \item Depth: Conceptually based on frequency domain analysis.
        \item Parameters: Varies depending on specific implementation but generally limited.
        \item Image Input Size: Adaptive to the image dimensions.
    \end{itemize}
    \item Incremental Coding Length (ICL) Model (Information-Theoretic) 
    \begin{itemize}
        \item Depth: The depth concept doesn't directly apply; it's based on information theory principles.
        \item Parameters: Typically small, as it deals with information measures.
        \item Image Input Size: Adaptive to the image dimensions.
    \end{itemize}
    \item DCTS Model (Spectral-Oriented) 
    \begin{itemize}
        \item Depth: Depth is not a primary consideration; it focuses on spectral features.
        \item Parameters: Generally limited, as it extracts spectral information.
        \item Image Input Size: Flexible, depending on the specific spectral analysis method.
    \end{itemize}
\end{enumerate}

The summary of the above model is presented in Table.\ref{fig:model_infos}.

\begin{table}[H]
\begin{center}
\caption{\textbf{Model summary information.} The pretrained networks and psychophysical models and some of their properties (e.g., depth, parameters and image input size).}
\resizebox{\columnwidth}{!}{%
\begin{tabular}{||c c c c||} 
 \hline
 Network & Depth & Parameters (Millions) & Image Input Size \\ [0.3ex] 
 \hline\hline
 Alexnet & 8 & 61.0	& 227-by-227 \\ 
 \hline
 Googlenet & 22 & 7.0	& 224-by-224  \\
 \hline
 Inceptionv3 & 48 & 23.9 & 299-by-299  \\
 \hline
 Densenet201 & 201 & 20.0 & 224-by-224 \\
 \hline
 DeepGazeII & 19+ & - & - \\
 \hline
 HFT & - & - & - \\
 \hline
 ICL & - & - & - \\
 \hline
 DCTS & - & - & - \\
 \hline
\end{tabular}
}
\label{fig:model_infos}
\end{center}
\end{table}

\subsection{Pre-trained CNNs and psychophysical models on psychophysical images}
\label{ssec:models}

Indeed, as previously mentioned, our investigation involved the utilization of pre-trained neural networks, including AlexNet, GoogLeNet, Inceptionv3, Densenet201, and DeepGazeII, for the purpose of visualizing saliency maps when using psychophysical synthetic images.

\begin{figure}[htb]
\begin{minipage}[b]{1.0\linewidth}
  \centering
  \centerline{\includegraphics[width=9cm, height=13em]{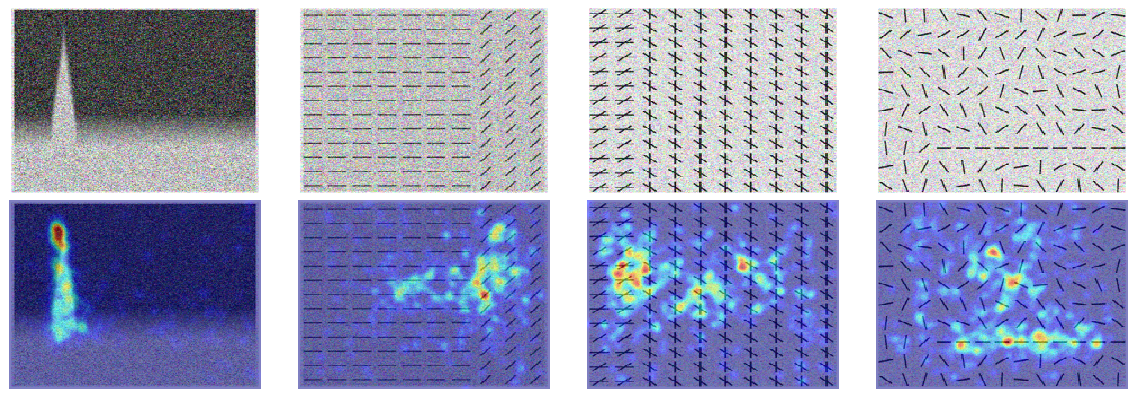}}
\end{minipage}
\caption{\textbf{A specific portion of the degraded psychophysical images was chosen for analysis.} The top row displays psychophysical stimulus images that have been intentionally degraded, while the second row showcases saliency prediction maps generated by human observers.}
\label{fig:sid4vam_s_noise1}
\end{figure}
\begin{figure*}[ht]
\centering
\centerline{\includegraphics[width=0.95\textwidth, height=33em]{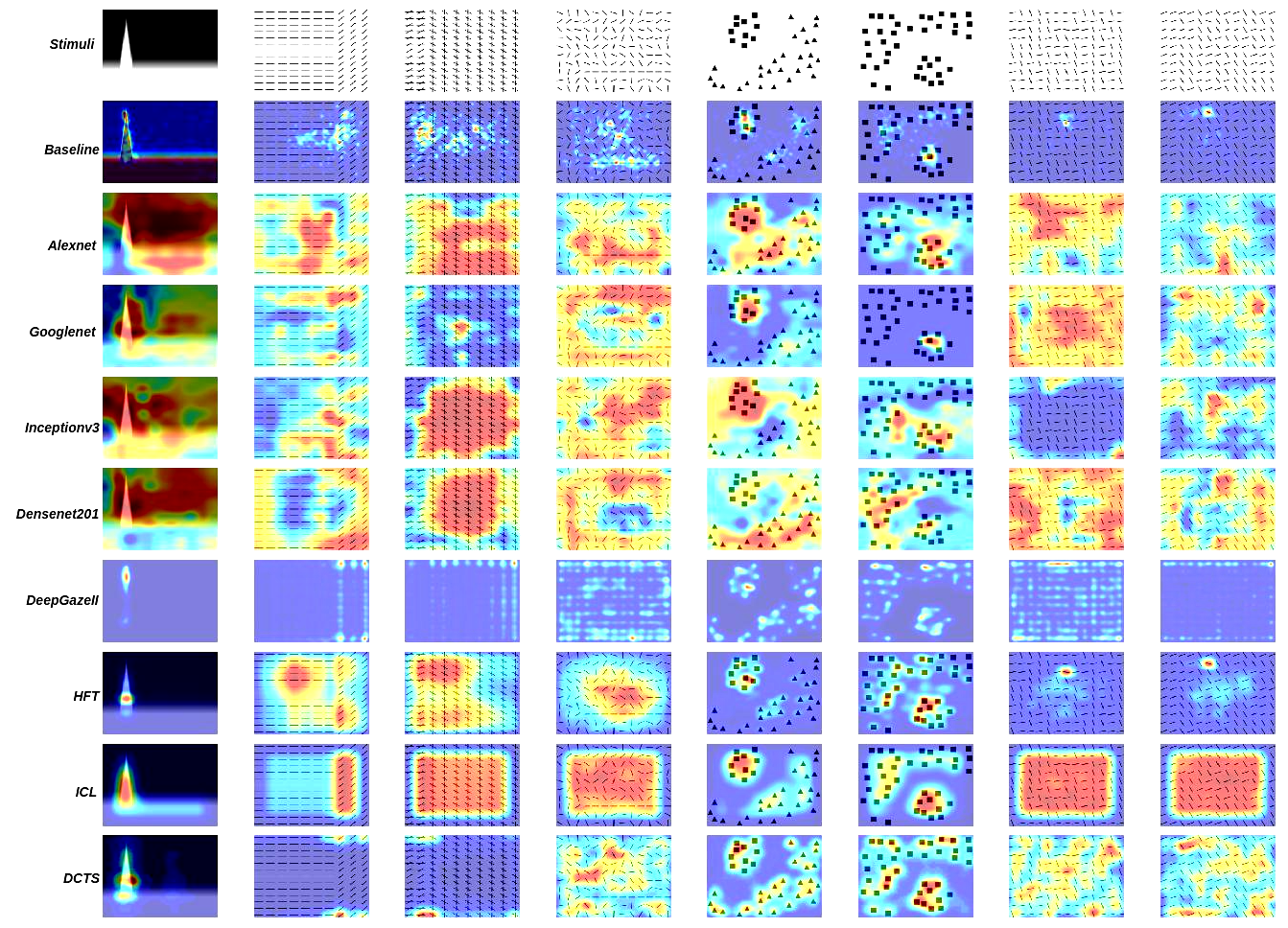}}
\caption{\textbf{Predicting saliency using pre-trained convolutional neural networks and biologically inspired saliency prediction models on selected images.} The top row shows psychophysical stimuli images, followed by the second row, which shows the saliency prediction map from humans. The remaining rows show saliency prediction maps from different models.}
\label{fig:sid4vam_s_from_pretrainedCNNs}
\end{figure*}

In parallel, we incorporated classical Fourier and spectral saliency prediction models, specifically the HTF, ICL, and DCTS models, which draw inspiration from low-level human vision functions. In the case of pre-trained neural networks, saliency maps were computed using the gradient-weighted class activation mapping (Grad-CAM) technique. Grad-CAM is a straightforward method employed to gain insights into which regions of an image hold the greatest significance for recognition by deep neural networks \cite{SelvarajuCDVPB17}.

Use Grad-CAM to gain a high-level understanding of what image features a network uses to make a particular classification or perform other tasks. The Grad-CAM map for a convolutional layer with $k$ feature maps (channels), $A_{i, j}^{k}$, could be estimated with,

\begin{equation}
S=\operatorname{ReLU}\left(\sum_{k} \frac{1}{N} \sum_{i} \sum_{j} \frac{\partial y^{c}}{\partial A_{i, j}^{k}} A^{k}\right)
\end{equation}
 
where $y^{c}$ is output, representing the score for class $c$, $i,j$ indexes the pixels, $N$ is the total number of pixels in the feature map. The rectified linear unit (ReLU) activation ensures you get only the features that have a positive contribution to the class of interest. The output is therefore a heatmap for the specified class. The HTF, ICL, and DCTS models were designed with their foundations rooted in Fourier-spectral and information-theoretic principles, respectively. For comprehensive details about these models, please consult Table~\ref{fig:model_infos}.

\subsection{Pre-trained CNNs and psychophysical models on degraded psychophysical images}
\label{ssec:demodels}
In our study, we aimed to assess the robustness of different saliency prediction models when confronted with noise interference, as illustrated in Fig.~\ref{fig:sid4vam_s_noise1}. To introduce noise, we applied salt-and-pepper noise with a density of 0.3 to each of the psychophysical images. Subsequently, we supplied these degraded images as inputs to various models to investigate the impact of noise on the accuracy of both CNNs and psychophysical models in the context of saliency prediction.

\begin{figure*}[ht!]
\begin{minipage}[b]{1.0\linewidth}
  \centering
  \centerline{\includegraphics[width=15cm, height=15em]{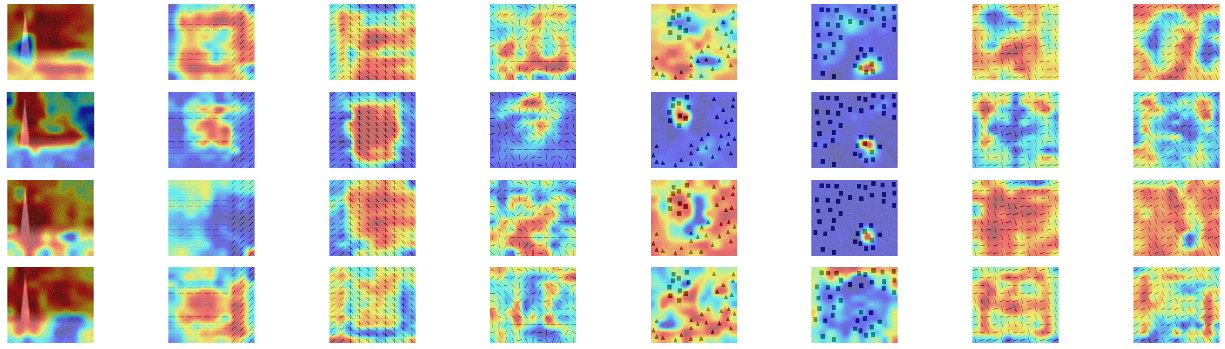}}
  \textcolor{black}{\rule{16cm}{1mm}}
  \centerline{\includegraphics[width=0.85\textwidth, height=18em]{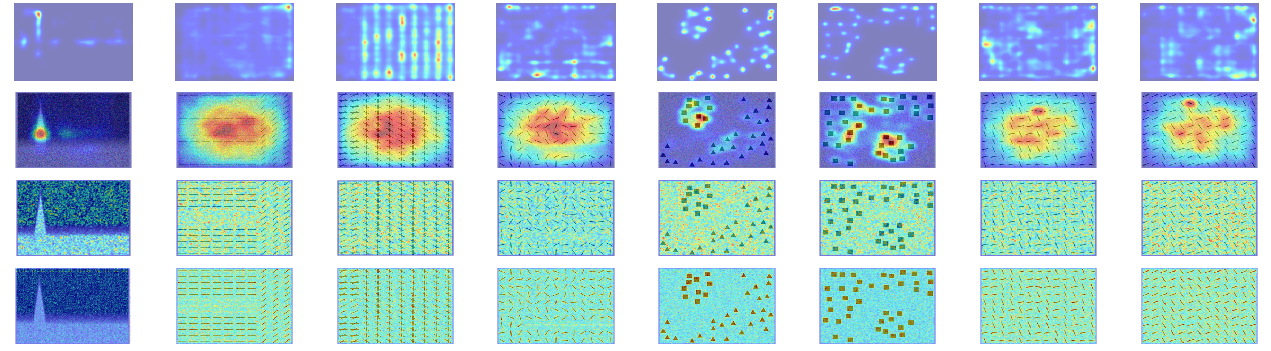}}
\end{minipage}
\caption{\textbf{Saliency predictions were made on selected degraded images using pre-trained convolutional neural networks and psychophysical models.} The presentation includes four rows at the top displaying degraded saliency predictions generated by pre-trained CNNs. The subsequent rows exhibit saliency prediction maps derived from psychophysical models. Specifically, from the top row to the bottom row, the saliency prediction results are ordered as follows: AlexNet, GoogLeNet, Inceptionv3, Densenet201, HTF, ICL, and DCTS.}
\label{fig:sid4vam_s_noise2}
\end{figure*}

\subsection{Evaluation metrics}
\label{ssec:metrics}
We employed accuracy (ACC) as a key statistical measure to evaluate the performance of saliency prediction. ACC is defined as follows:

\begin{equation}
    \mathrm{ACC}=\frac{\mathrm{TP}+\mathrm{TN}}{\mathrm{P}+\mathrm{N}}
\end{equation}

where TP, TN, P, and N refer to true positive, true negative, condition positive, and condition negative, respectively. In this study, TP is the number of positive samples correctly predicted as positive, TN the number of correctly predicted negative samples, P the total number of positive samples, and N the total number of negative samples.

\section{Results}
\label{sec:resu}
The visualization of saliency predictions generated by the aforementioned models is depicted in Fig.~\ref{fig:sid4vam_s_from_pretrainedCNNs}. In this figure, we have plotted saliency prediction maps produced by different models. Upon careful observation, it becomes evident that the DCTS and HFT models exhibit saliency predictions that closely resemble human-level predictions, as indicated by the baseline in the second row. However, when compared to the psychophysical models, the pre-trained CNNs yield less accurate results.

\begin{figure}[ht!]
\begin{minipage}[b]{1.0\linewidth}
  \centering
  \centerline{\includegraphics[width=\textwidth, height=18em]{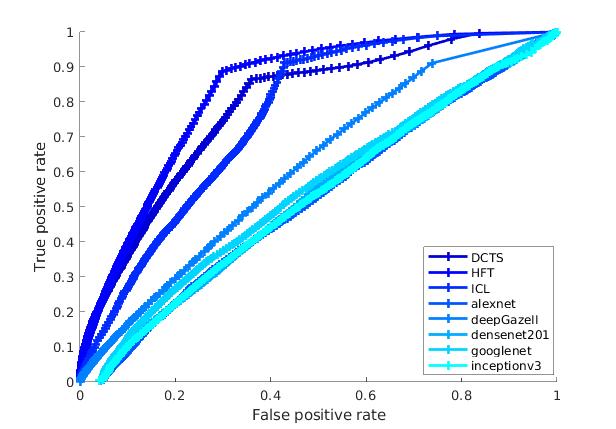}}
\end{minipage} 
\caption{\textbf{AUC - ROC curve.} The performance of saliency prediction was assessed using pre-trained convolutional neural networks (CNNs) with varying depths and biologically inspired saliency prediction models.}
\label{fig:sid4vam_s_metric}
\end{figure}

To perform a statistical analysis of the saliency prediction performance with different models, we utilized the receiver operator characteristic (ROC) curve, as depicted in Fig.~\ref{fig:sid4vam_s_metric}. The statistical results affirm that psychophysical models outperform pre-trained CNNs in terms of prediction accuracy. The reasons for these observations will be elucidated in the subsequent section.

Furthermore, we explored how various models are influenced by noise interference when operating in a noisy environment. In Fig.~\ref{fig:sid4vam_s_noise2}, we made the discovery that for pre-trained CNNs (AlexNet, GoogLeNet, Inceptionv3, and Densenet201), and even when using purpose-built neural networks like DeepGazeII, the addition of noise had minimal impact on their performance. Conversely, psychophysical models (HFT, ICL, and DCTS) were considerably affected by the introduction of noise. This result aligns with expectations since pre-trained neural networks come equipped with fixed parameters and weights tailored for natural images. In contrast, psychophysical models require the input of degraded images without a learning process to produce saliency maps.

It's important to note that the evaluation of saliency performance on degraded images is currently constrained by the absence of human saliency prediction data for such images. Consequently, we solely report saliency prediction metrics for clean images in Fig.~\ref{fig:sid4vam_s_metric}. However, future research endeavors could involve the collection of human saliency prediction data for degraded images, which would present a compelling avenue for exploration.

\section{Conclusion and Future work}
\label{sec:conclu}
This study delved into the accuracy of saliency prediction using CNNs, Fourier, and spectral models inspired by low-level vision systems. On psychophysical synthetic images, it was observed that saliency prediction models inspired by low-level human vision functions outperformed pre-trained deep neural networks. These findings underscored the fact that CNNs do not precisely emulate the workings of the human eye's low-level vision system. It is important to exercise caution when applying deep CNNs trained on natural images to tasks involving psychophysical synthetic images, as this may lead to suboptimal results. Additionally, the study revealed that psychophysical models are more susceptible to the effects of noise when compared to pre-trained neural networks.

Several challenges were identified when utilizing psychophysical images with CNNs. First, there is often an inadequate quantity of psychophysical images available for training deep CNNs. Furthermore, CNN architectures primarily model high-level human visual functions, but the extent to which they replicate low-level visual functions such as color, orientation, brightness adjustments, and adaptation remains a subject of investigation. Additionally, there is room for expanding research on saliency prediction for degradation distraction tasks, encompassing a wider range of image degradations for both clean natural images and psychophysical synthetic images, in order to assess saliency prediction stability. Collecting human saliency prediction data for degraded images as a benchmark for model performance is essential and should be considered for future studies.

Lastly, it is imperative to emphasize the significance of incorporating psychophysical methods when examining the properties of artificial neural networks. Visual neuroscientists have conducted numerous psychophysical studies on visual attention mechanisms, significantly advancing our understanding of these domains. Integrating these insights into artificial neural networks can enhance their generality and stability. However, this area remains underexplored, and there is a need for further research in this direction. Incorporating additional psychophysical studies into the development and analysis of artificial neural networks holds the potential to deepen our comprehension of both the human visual system and the mechanisms of artificial neural networks.



\bibliographystyle{IEEEbib}
\bibliography{strings}

\begin{thebibliography}{10}

\bibitem{Ashish17}
Ashish Vaswani, Noam Shazeer, Niki Parmar, Jakob Uszkoreit, Llion Jones,
  Aidan~N Gomez, \L~ukasz Kaiser, and Illia Polosukhin,
\newblock ``Attention is all you need,''
\newblock in {\em Advances in Neural Information Processing Systems}, I.~Guyon,
  U.~V. Luxburg, S.~Bengio, H.~Wallach, R.~Fergus, S.~Vishwanathan, and
  R.~Garnett, Eds. 2017, vol.~30, Curran Associates, Inc.

\bibitem{PengHZ18}
Yuxin Peng, Xiangteng He, and Junjie Zhao,
\newblock ``Object-part attention model for fine-grained image
  classification.,''
\newblock {\em IEEE Trans. Image Process.}, vol. 27, no. 3, pp. 1487--1500,
  2018.

\bibitem{Carrasco11}
Marisa Carrasco,
\newblock ``Visual attention: The past 25 years,''
\newblock {\em Vision Research}, vol. 51, no. 13, pp. 1484--1525, 2011,
\newblock Vision Research 50th Anniversary Issue: Part 2.

\bibitem{QiangNN22}
Qiang Li,
\newblock ``Functional connectivity inference from fmri data using multivariate
  information measures,''
\newblock {\em Neural Networks}, vol. 146, pp. 85--97, 2022.

\bibitem{Zhaoping06}
Zhaoping Li,
\newblock ``Theoretical understanding of the early visual processes by data
  compression and data selection,''
\newblock {\em Network: Computation in Neural Systems}, vol. 17, no. 4, pp.
  301--334, 2006,
\newblock PMID: 17283516.

\bibitem{Kelly62}
Donald~H. Kelly,
\newblock ``Information capacity of a single retinal channel,''
\newblock {\em IRE Transactions on Information Theory}, vol. 8, no. 3, pp.
  221--226, 1962.

\bibitem{George56}
George~C. Sziklai,
\newblock ``Some studies in the speed of visual perception,''
\newblock {\em IRE Transactions on Information Theory}, vol. 2, no. 3, pp.
  125--128, 1956.

\bibitem{Barlow61}
Horace Barlow,
\newblock ``Possible principles underlying the transformations of sensory
  messages,''
\newblock {\em Sensory Communication}, vol. 1, 01 1961.

\bibitem{Itti98}
Laurent Itti, Christof Koch, and Ernst Niebur,
\newblock ``A model of saliency-based visual attention for rapid scene
  analysis,''
\newblock {\em IEEE Transactions on Pattern Analysis and Machine Intelligence},
  vol. 20, no. 11, pp. 1254--1259, 1998.

\bibitem{Borji2021SaliencyPI}
Ali Borji,
\newblock ``Saliency prediction in the deep learning era: Successes and
  limitations,''
\newblock {\em IEEE Transactions on Pattern Analysis and Machine Intelligence},
  vol. 43, pp. 679--700, 2021.

\bibitem{Cornia16}
Marcella Cornia, Lorenzo Baraldi, Giuseppe Serra, and Rita Cucchiara,
\newblock ``A deep multi-level network for saliency prediction,''
\newblock in {\em 2016 23rd International Conference on Pattern Recognition
  (ICPR)}, 2016, pp. 3488--3493.

\bibitem{Kmmerer2016}
Matthias K{\"u}mmerer, Thomas S.~A. Wallis, and Matthias Bethge,
\newblock ``Deepgaze ii: Reading fixations from deep features trained on object
  recognition,''
\newblock {\em ArXiv}, vol. abs/1610.01563, 2016.

\bibitem{NIPS2012_4824}
Alex Krizhevsky, Ilya Sutskever, and Geoffrey~E. Hinton,
\newblock ``Imagenet classification with deep convolutional neural networks,''
\newblock in {\em Advances in Neural Information Processing Systems 25},
  F.~Pereira, C.~J.~C. Burges, L.~Bottou, and K.~Q. Weinberger, Eds., pp.
  1097--1105. Curran Associates, Inc., 2012.

\bibitem{szegedy2014going}
Christian Szegedy, Wei Liu, Yangqing Jia, Pierre Sermanet, Scott Reed, Dragomir
  Anguelov, Dumitru Erhan, Vincent Vanhoucke, and Andrew Rabinovich,
\newblock ``Going deeper with convolutions,''
\newblock {\em IEEE Conference on Computer Vision and Pattern Recognition
  (CVPR)}, pp. 1--9, 06 2015.

\bibitem{incep}
Christian Szegedy, Vincent Vanhoucke, Sergey Ioffe, Jonathon Shlens, and
  Zbigniew Wojna,
\newblock ``Rethinking the inception architecture for computer vision,''
\newblock {\em IEEE Conference on Computer Vision and Pattern Recognition
  (CVPR)}, vol. abs/1512.00567, 2015.

\bibitem{Huang2017DenselyCC}
Gao Huang, Zhuang Liu, and Kilian~Q. Weinberger,
\newblock ``Densely connected convolutional networks,''
\newblock {\em IEEE Conference on Computer Vision and Pattern Recognition
  (CVPR)}, pp. 2261--2269, 2017.

\bibitem{QiangLI20}
Qiang Li,
\newblock ``Saliency prediction based on multi-channel models of visual
  processing,''
\newblock {\em Machine Vision and Applications}, vol. 34, 05 2023.

\bibitem{Jian12}
Jian li, Martin Levine, Xiangjing An, Xin Xu, and Hangen He,
\newblock ``Visual saliency based on scale-space analysis in the frequency
  domain,''
\newblock {\em IEEE Transactions on Pattern Analysis and Machine Intelligence},
  vol. 35, pp. 996--1010, 11 2012.

\bibitem{Hou08}
Xiaodi Hou and Liqing Zhang,
\newblock ``Dynamic visual attention: Searching for coding length increments,''
\newblock in {\em Adv. Neural Inf. Process. Syst}, 01 2008, vol.~21, pp.
  681--688.

\bibitem{Hou11}
Xiaodi Hou, Jonathan Harel, and Christof Koch,
\newblock ``Image signature: Highlighting sparse salient regions,''
\newblock {\em IEEE transactions on pattern analysis and machine intelligence},
  vol. 34, 07 2011.

\bibitem{Berga2019SID4VAMAB}
David Berga, Xos{\'e}~Ram{\'o}n Fdez-Vidal, Xavier Otazu, and Xose~Manuel
  Pardo,
\newblock ``Sid4vam: A benchmark dataset with synthetic images for visual
  attention modeling,''
\newblock {\em IEEE/CVF International Conference on Computer Vision (ICCV)},
  pp. 8788--8797, 2019.

\bibitem{BERGA201960}
David Berga, Xosé~R. Fdez-Vidal, Xavier Otazu, Víctor Leborán, and Xosé~M.
  Pardo,
\newblock ``Psychophysical evaluation of individual low-level feature
  influences on visual attention,''
\newblock {\em Vision Research}, vol. 154, pp. 60--79, 2019.

\bibitem{SelvarajuCDVPB17}
Ramprasaath~R. Selvaraju, Michael Cogswell, Abhishek Das, Ramakrishna Vedantam,
  Devi Parikh, and Dhruv Batra,
\newblock ``Grad-cam: Visual explanations from deep networks via gradient-based
  localization.,''
\newblock in {\em ICCV}. 2017, pp. 618--626, IEEE Computer Society.

\end{thebibliography}
\end{document}